\documentclass[preprint,preprintnumbers,amsmath,amssymb]{revtex4}

\usepackage{graphicx}
\usepackage{dcolumn}
\usepackage{bm}
\usepackage{epsfig}
\usepackage{color}

\begin{document}


\title{Supplemental Material: Eshelby ensemble of highly viscous flow out of equilibrium}

\author{U. Buchenau}

\affiliation{%
Forschungszentrum J\"ulich GmbH, J\"ulich Centre for Neutron Science (JCNS-1) and Institute for Complex Systems (ICS-1),  52425 J\"ulich, GERMANY
}%

\date{May 12, 2019}
\maketitle

\section{Evidence on the size of the Eshelby domains}

Direct evidence on the size of the rearranging regions is found in the relaxation spectrum of a metallic glass at room temperature \cite{atzmon}. The time dependence of the shear relaxation shows weak wiggles with an inter-wiggle distance corresponding to a barrier difference of 2 k$_B$T at room temperature, a bit less than k$_B$T at the glass temperature of about 640 K of the alloy. The most obvious interpretation is an increase by adding one atom to the Eshelby region from barrier to barrier. Since the barrier at the glass transition is about 35 k$_B$T, this implies a number $N_c$ of about forty atoms in the Eshelby region responsible for the viscous flow.

The value of forty atoms in metallic glasses is consistent with the following estimate \cite{asyth1}: For a transforming Eshelby domain consisting of $N$ particles, the Kohlrausch $\beta$ results from the increase of the number of structural possibilities with increasing $N$, combined with the increasing barrier height with increasing $N$. Let $S_1$ be the structural entropy per particle, $V_1$ the increase of the barrier $V_c$ per particle. Then
\begin{equation}\label{v1}
	\beta=\frac{S_1T}{V_1}, 
\end{equation}
so for a barrier height of about 35 k$_B$T at $T_g$
\begin{equation}\label{N}
	N_c=\frac{35\beta k_B}{S_1}.
\end{equation}
In metallic glasses \cite{met}, the Kohlrausch $\beta$ is almost always 0.4. $S_1$ at $T_g$ is 0.36 k$_B$ per atom in Pd$_{40}$Ni$_{40}$P$_{20}$ \cite{wildek} and in vitralloy-1 \cite{busch}, in agreement with forty atoms from eq. (\ref{N}).

Eq. (\ref{v1}) predicts $V_1=0.036$ eV for selenium, with $\beta=0.31$ \cite{roland} and $S_1$=0.43 k$_B$ per atom \cite{chang}. This prediction is consistent with wiggles in the ultrasound absorption in the glass phase of selenium \cite{dangelo}.

In principle, the determination of $S_1$ requires dedicated heat capacity measurements of both phases, undercooled liquid and crystal. However, if one has only heat capacity measurements of the undercooled liquid at $T_g$, together with a Vogel-Fulcher temperature $T_{VF}$ from dynamical measurements, one can make use of the Adam-Gibbs relation in the form proposed by Richert and Angell \cite{ra}
\begin{equation}
	S_{exc}=S_\infty\left(1-\frac{T_{VF}}{T}\right)
\end{equation}
which has been found to describe the excess entropy of several molecular glass formers fairly well, at least close to $T_g$.

If this holds, one can determine $S_{exc}$ at $T_g$ from the heat capacity difference $\Delta c_p$ between undercooled liquid and glass via
\begin{equation}\label{s1}
	S_{exc}(T_g)=\Delta c_p\left(\frac{T_g}{T_{VF}}-1\right).
\end{equation}

In polystyrene with $\Delta c_p=0.235$ J/gK = 2.94 k$_B$ per monomer \cite{hensel} and $T_{VF}=319$ K (see Table I of the main paper), this yields $S_1=0.475$ k$_B$ per monomer, which for $\beta$=0.33 \cite{plazek} gives an $N_c$ of 23.6 monomers. 

\begin{figure}    
\hspace{-0cm} \vspace{0cm} \epsfig{file=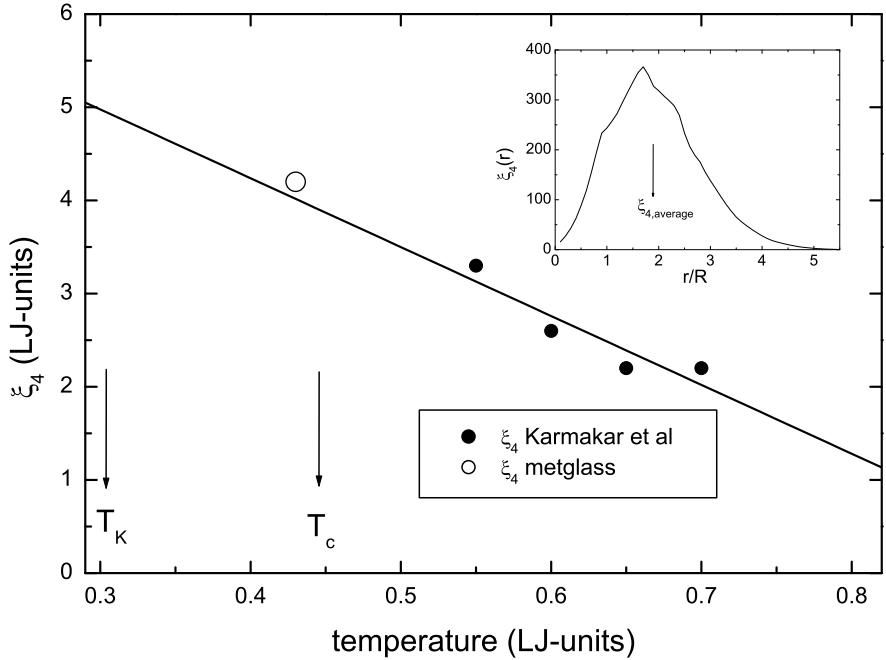,width=12 cm,angle=0} \vspace{0cm} \caption{Four-point correlation lengths in the binary Lennard-Jones system \cite{karmakar}, extrapolated down to the value calculated for a metallic glass at its glass temperature from shear relaxation data \cite{atzmon}. $T_K$ is the Kauzmann temperature, $T_c$ the critical temperature of the mode coupling theory. The insert shows the calculated four-point correlation function for the Eshelby shear transformation of a sphere with radius $R$.}
\end{figure}

The number of forty atoms in the metallic glasses is consistent with numerical evidence. Fig. 1 shows the temperature dependence of the four-point correlation length \cite{karmakar} calculated for the binary Lennard-Jones system.

It is straightforward to calculate the four-point correlation length for an Eshelby shear transformation like the one in Fig. 1 of the main paper for a sphere. Let $\vec{r}\pm\vec{u}/2$ be the coordinates of the volume $\delta V$ before and after the transformation. Then $u_1^2u_2^2\delta V_1\delta V_2/16$ is the contribution of the two volumes $\delta V_1$ and $\delta V_2$ to the four-point amplitude correlation function $f_4(r)=<u_1(0)u_1(t)u_2(0)u_2(t)>$ at the distance $r$ between $\vec{r}_1$ and $\vec{r}_2$, assuming that the transformation occurred between 0 and $t$. Integrating this for all point pairs in the volume, one gets a function which peaks close to its mean value 1.92 $R$, where $R$ is the radius of the sphere. In this calculation, the displacement field outside the sphere, which is not easy to calculate \cite{eshelby}, was simplified to an $(r/R)^2$ decay of the amplitude at the corresponding point on the sphere surface. The function is plotted in the insert of Fig. 1.

To translate $R$ into Lennard-Jones units, remember that the nearest neighbor distance is $2^{1/6}=1.12$ Lennard-Jones units and that the packing of spheres in disorder fills the volume by 66 percent. Taking the nearest neighbor distance $d_{at}$ as the diameter of the atomic spheres, one has
\begin{equation}
	\frac{4\pi^2}{3}R^3=\frac{40}{0.66}\frac{4\pi^2}{3}\frac{d_{at}^3}{8},
\end{equation}
so $R=1.96d_{at}$, which brings $\xi_4$ to 3.87 Lennard-Jones units.

In order to determine the appropriate Lennard-Jones temperature for the metallic glass, one can identify the Vogel-Fulcher temperature of the metallic glass former with the Kauzmann temperature of the binary Lennard-Jones system. For Pd$_{40}$Ni$_{40}$P$_{20}$ \cite{wildek}, this places the glass temperature at 0.43 Lennard-Jones units.

The whole consideration is further supported by evidence for Eshelby domains in numerical work on two-dimensional Lennard-Jones systems \cite{lemaitre}.

But there are also results \cite{biroli} which do not fit so well into this picture, namely point-to-set correlation lengths for the binary Lennard-Jones system which are a factor of 1.5 smaller than the four-point correlation values in Fig. 1. This implies that one has to enclose the Eshelby sphere into a rather small rigid sphere, with a radius only 1.77/1.5=1.18 times larger than the one of the Eshelby sphere itself, before it stops moving. This ratio seems rather small.

On the other hand, it was necessary to use very large samples to obtain the values in Fig. 1. If one takes smaller samples, like those used in the determination of the point-to-set values \cite{biroli}, one gets four-point correlation values which are also a factor of 1.5 smaller \cite{karmacomm}. Possibly, the point-to-set values (and the $\lambda_{min}$-values determined from the boson peak modes which coincide with them \cite{biroli}) would also increase for larger samples.

\section{The three other systems of Hensel and Schick}

\begin{figure}   
\hspace{-0cm} \vspace{0cm} \epsfig{file=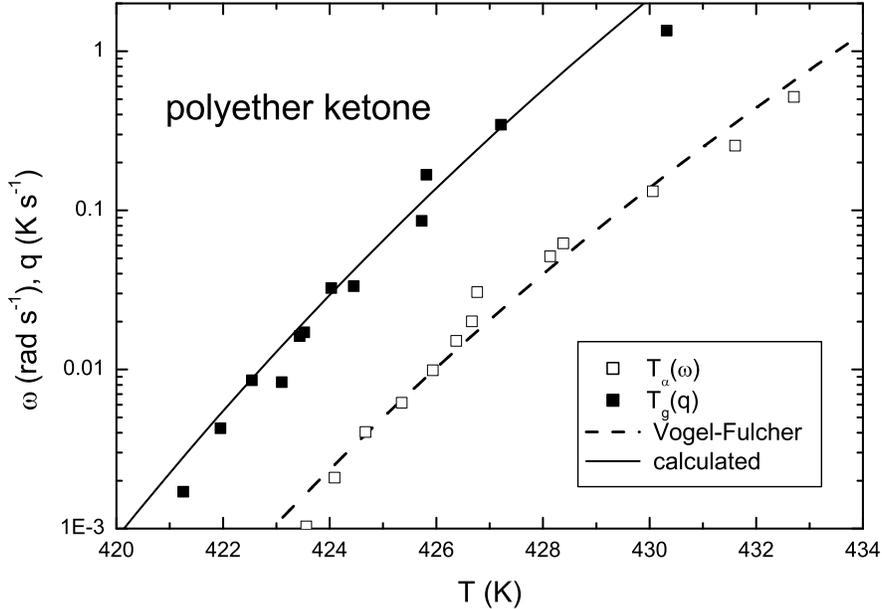,width=12 cm,angle=0} \vspace{0cm} \caption{$T_g(q)$ and $T_\alpha(\omega)$ for different cooling rates and frequencies in polyether ketone \cite{hensel} (parameters see Table I of the main paper).}
\end{figure}

\begin{figure}   
\hspace{-0cm} \vspace{0cm} \epsfig{file=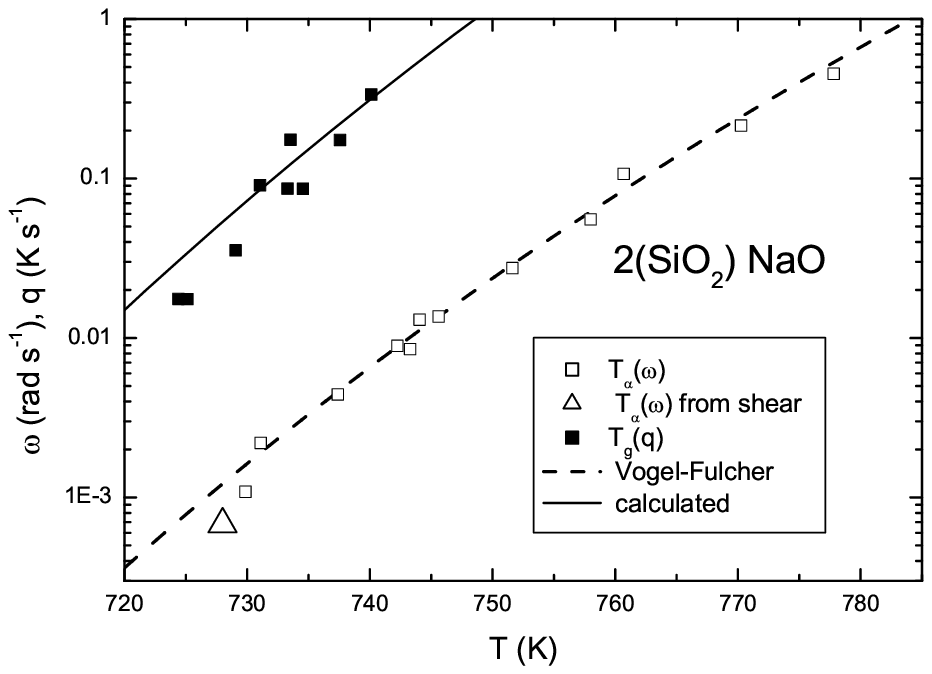,width=12 cm,angle=0} \vspace{0cm} \caption{$T_g(q)$ and $T_\alpha(\omega)$ for different cooling rates and frequencies in 2(SiO$_2$)NaO \cite{hensel} (parameters see Table I). The triangle is $T_\alpha(\omega)$ from a dynamical shear measurement \cite{mills}.}
\end{figure}

\begin{figure}   
\hspace{-0cm} \vspace{0cm} \epsfig{file=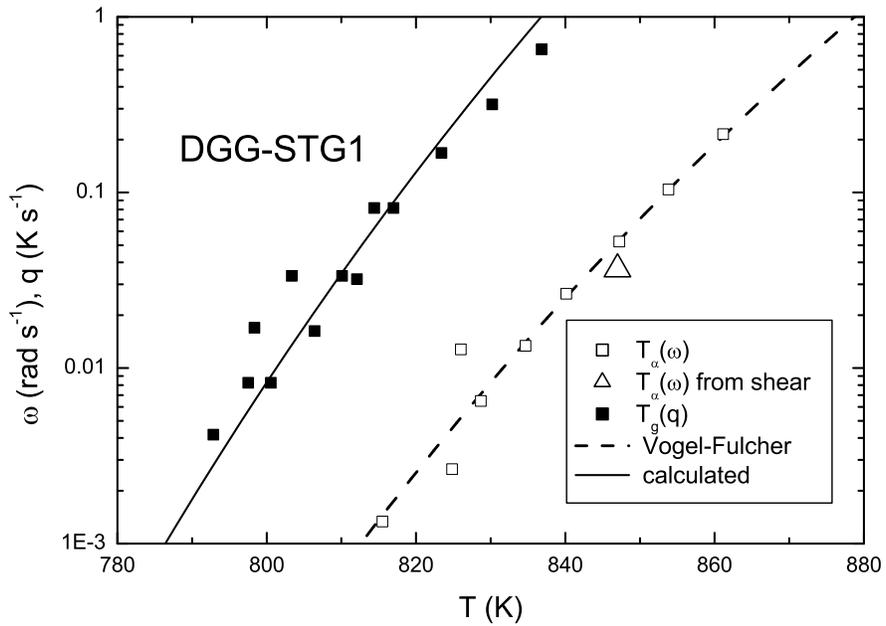,width=12 cm,angle=0} \vspace{0cm} \caption{$T_g(q)$ and $T_\alpha(\omega)$ for different cooling rates and frequencies in the window glass DGG-STG1 \cite{hensel} (parameters see Table I).  The triangle is $T_\alpha(\omega)$ from a dynamical shear measurement \cite{donth2}.}
\end{figure}

Figs. 2 to 4 show the comparison of equilibrium $T_\alpha(\omega)$ and non-equilibrium $T_g(q)$ data of Hensel and Schick \cite{hensel}, together with their fits, the equilibrium data providing the two Vogel-Fulcher parameters in Table I of the main paper, which allow to calculate the curve for the non-equilibrium data.

Polyether ketone in Fig. 2 is a polymer, where eq. (4) of the main paper does not apply. But for the two silicate glass formers in Fig. 3 and Fig. 4, it is found to be valid within the error bars to be expected for measurements in different furnaces at different laboratories. The triangle for $T_\alpha(\omega)$ in Fig. 3 is calculated from equs. (4) and (6) of the main paper, with $\eta$ and $G$ taken from shear data \cite{mills} of 2(SiO$_2$)NaO at 728 K, the one in Fig. 4 from shear data \cite{donth2} of the window glass.   

\section{Separation of segmental relaxation and Rouse relaxation in polystyrene}

\begin{figure}   
\hspace{-0cm} \vspace{0cm} \epsfig{file=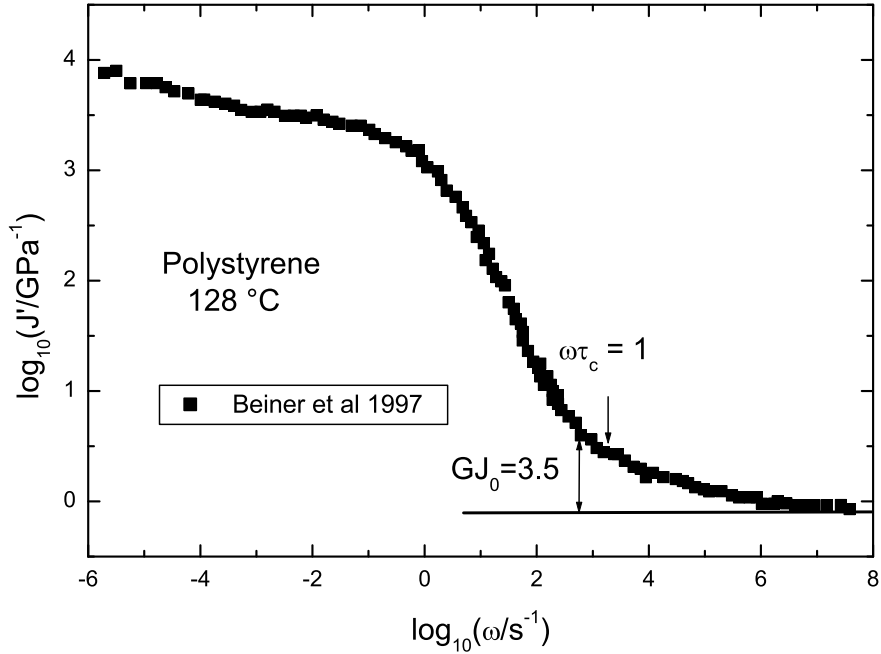,width=12 cm,angle=0} \vspace{0cm} \caption{Master curve for the shear compliance $J'(\omega)$ of polystyrene at 128 degrees Celsius \cite{beiner}, showing that the crossover from segmental relaxation to Rouse modes occurs close to $\omega\tau_c=1$, with $\tau_c$ calculated from the Vogel-Fulcher parameters of Table I of the main paper.}
\end{figure}

Polymers differ from other glass formers in the recoverable shear compliance $J_0$, which one measures after removing the shear stress of a stationary flow experiment. In non-polymeric glass formers, one finds values $GJ_0$ between 2 and 3, implying a relaxational back-jump contribution $GJ_{0,rel}$ between 1 and 2 ($J_0$ naturally contains the elastic contribution $1/G$)  \cite{plazek-bero}. The irreversible Eshelby explanation of the viscosity \cite{asyth1} attributes these back-jumps to reversible Eshelby transitions of smaller domains, at times shorter than $\tau_c$, responsible for the Kohlrausch short time shear and dielectric response proportional to $t^\beta$ with $\beta\approx1/2$.

A polymer has a much larger recoverable compliance \cite{plazek,roland} than a simple glass former, because the Rouse modes also lead to a recoverable compliance. But the Rouse modes are irreversible in the same sense as the irreversible Eshelby rearrangements (in fact, they should consist out of many irreversible Eshelby transitions); though they lead back to a chain configuration of approximately the same shear strain, they do not lead back to exactly the same chain configuration. Thus in a polymer one has to distinguish between reversible recoverable compliance contributions at relaxation times shorter than $\tau_c$ and irreversible ones at relaxation times much longer than $\tau_c$.

The analysis of polystyrene mechanical data separating segmental relaxation and Rouse modes \cite{ngai} arrives at $GJ_0\approx 3.5$ for the segmental relaxation. Indeed, the condition $GJ(\omega)=3.5$ for polystyrene \cite{beiner} lies close to $\omega\tau_c=1$, with $\tau_c$ calculated from the Vogel-Fulcher parameters in Table I. This is shown in Fig. 5.

\end{document}